\documentclass[twocolumn,showpacs,amsmath,amssymb,prl]{revtex4}

\usepackage{graphicx}
\usepackage{dcolumn}
\usepackage{bm}

\begin{document}

\title{Continuum Description of the Cytoskeleton: Ring Formation in the Cell Cortex}

\author{Alexander Zumdieck$^{1}$}
\thanks{These authors contributed equally to this work.}
\author{Marco Cosentino Lagomarsino$^{2,*}$}
\altaffiliation[Current address:]{Institut Curie,Paris,France}
\author{Catalin Tanase$^{2,*}$}
\altaffiliation[Current address:]{Institute of Theoretical Physics,
Utrecht University, The Netherlands}
\author{Karsten Kruse$^{1}$}
\author{Bela Mulder$^{2}$}
\author{Marileen Dogterom$^{2}$}
\author{Frank J\"{u}licher$^{1}$}
\affiliation{$^{1}$Max-Planck-Institut f\"{u}r Physik komplexer Systeme, Dresden, Germany}
\affiliation{$^{2}$FOM Institute for Atomic and Molecular Physics (AMOLF), Amsterdam, The Netherlands}

\begin{abstract}
Motivated by the formation of ring-like filament structures  in the cortex of plant 
and animal cells, we study the dynamics of a two-dimensional layer of cytoskeletal 
filaments and motor proteins near a surface by a general continuum theory.  As a 
result of active processes, dynamic patterns of filament orientation and density emerge via instabilities. We show that self-organization phenomena can lead to the formation of stationary and oscillating rings. 
We present state diagrams which reveal a rich scenario of asymptotic 
behaviors and discuss the role of boundary conditions.
 \end{abstract}


\maketitle

The cytoskeleton, an organized network of filamentous proteins, is an
essential component of all eukaryotic cells. It plays a major role in
morphogenesis, transport, motility and cell division~\cite{MBC}.
The key components of the cytoskeleton are actin filaments and
microtubules which are long elastic protein filaments that reach
lengths of several $\mu$m. They have 
chemically and dynamically distinct
ends and hence are structurally polar. 
Motor proteins consume chemical fuel to generate directed motion on these filaments and hence can induce stresses in the cytoskeletal network.

\begin{figure}
\includegraphics[scale=.4]{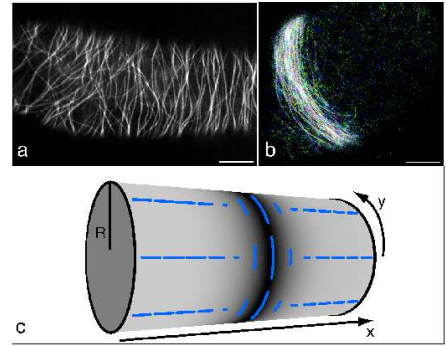}
\caption{\label{fig:fig1}
(color online)
Density and orientation of cortical filaments.
(a,b)
Fluorescence microscopy images of microtubule patterns in
Tobacco BY2 suspension cells. 
(a) Between cell divisions microtubules form a cortical array
and orient preferentially perpendicular to
the cell's long axis. (b) Shortly before cell division microtubules
condense into a ring-like structure, the Pre-Prophase Band (PPB).
(c) Representation of a stationary solution to the dynamic 
equations (\ref{eq:dcdt})-(\ref{eq:evolprojqxy}) resembling a PPB.
The density of filaments is represented by the grey level on the cylinder. The blue bars 
indicate the nematic order of filaments by their orientation and  length. 
(a,b) courtesy of Jan Vos, Wageningen University. Scale bars are 10\,$\mu$m.}
\end{figure}

An important cytoskeletal structure in various cell types are cortical rings
formed by bundles of filaments which wrap around the cell. Such rings
form within the cell cortex, a thin layer of filament network located close to the
cell membrane.
In animal cells a contractile ring containing actin filaments, cleaves the
dividing cell by generating a constriction~\cite{MBC}. In higher plant cells, microtubules
form cortical arrays (CA) between consecutive cell divisions~\cite{Was02}, with
a preferred orientation in the azimuthal direction, see Fig.~\ref{fig:fig1}(a). 
Shortly before entering division, the CA is dynamically reorganized 
and the preprophase band (PPB) is formed~\cite{Was02}. 
It consists of a ring-like bundle of microtubules 
that determines the location 
and orientation of the future division plane of the cell~\cite{GC00}, 
see Fig.~\ref{fig:fig1}(b). In addition to single rings, double rings have been
observed in certain cells~\cite{YAH+05}.

It has been experimentally observed  \emph{in vitro} that mixtures of
microtubules and motor proteins can self-organize  
into a number of patterns, such as bundles, 
asters and vortices~\cite{NSM+97,surr01}. Similar patterns where found in 
molecular dynamics simulations~\cite{surr01,Ned02}. 
Theoretical descriptions of cytoskeletal dynamics are either based on 
microscopic models for the motor-filament 
interactions~\cite{KJ00,BLJ00,KJ03,LM03} or consider 
the behavior on large length and time scales in terms of a continuum
theory~\cite{LK01,KPK+03,SMK04,KZJ03,krus04,ziebert:04}.
In contrast to microscopic descriptions, which rely on specific
and potentially incomplete microscopic pictures of the nonequilibrium
driving forces, a phenomenological description captures
generic conditions for the formation of patterns through instabilities. 
In the present work we develop a general coarse grained description of 
the two-diemensional cell cortex based on symmetry arguments and extend 
previous work to include nematic order. 

We focus on the
dynamical behaviour of a cortical layer of filaments 
at scales large compared to the filament lengths and the 
layer thickness, where a coarse grained, two-dimensional 
description is suitable. We assume that the state of
the system is essentially specified by the scalar mass density
$c(\bm{r},t)$, the polarization 
$\bm{p}(\bm{r},t)$, a vector along the
average direction
of the filaments, and $\bm{q}(\bm{r},t)$, a second
rank traceless symmetric tensor that measures orientational order.
These fields  represent slow hydrodynamic modes as
they are linked either to a conservation law (the density $c$) or to
a spontaneously broken symmetry (the fields $\bm{p}$ and $\bm{q}$).
Other degrees of freedom in the system are assumed to relax more rapidly
and are thus ignored \cite{krus04}.

The choice of the appropriate dynamic equations
for these fields is governed by the following principles 
(i) all dynamics is overdamped as its origin lies in the motion of colloidal sized
particles in a viscous background fluid, 
(ii) all mass transport is due to stresses generated in the system, and 
(iii) all driving terms are allowed that respect the rotational invariance of space.
This is consistent with the fact
that they are caused by local interactions between the microscopic
components. For the mass density we have the continuity equation
\begin{equation}
\partial _{t}c+\bm{\nabla} \cdot \bm{j}=s\quad ,  \label{eq:dcdt}
\end{equation}
where $\bm{j}$ is the filament current and $s$ represents source and sink
terms describing the polymerization and depolymerization of filaments. The
filament current is a consequence of inhomogeneities in the mechanical
stress in the filament system. 
For a layer of cortical filaments close to the cell surface 
moving against an immobile background fluid and assuming local 
isotropic friction, we thus write
\begin{equation}
\bm{j}=\eta ^{-1} \bm{\nabla} \cdot \bm{\sigma}\quad ,
\end{equation}
where $\bm{\sigma }$ is the stress tensor and $\eta $ an effective
friction coefficient. The dynamic equations for the
non-conserved fields $\bm{p}$ and $\bm{q}$ can be expressed as
$\partial _{t}\bm{p}= \bm{\psi}$,
and $\partial _{t}\bm{q}= \bm{\omega }$,
where $\bm{\psi}$ is a vector and $\bm{\omega }$ is a symmetric traceless tensor. 

These equations are closed by assuming that the
evolution of the system is completely determined by the present state of
the system. This implies that the stress tensor $\bm{\sigma}$ and the ``velocities'' of
the polarization and the nematic order parameter $\bm{\psi}$ and
$\bm{\omega }$ can be expressed in terms of the dynamic fields
$c$, $\bm{p}$ and $\bm{q}$. 
We then expand these driving forces
in terms of perturbations of the dynamic fields
and their spatial derivatives around a reference state that we assume to be homogenous. 

We describe the cell cortex as a two-dimensional active gel, embedded in
three dimensional space along the cell surface. 
Motivated by the morphology of plant cells, we wrap the 2d gel on a cylinder.
Experimental evidence indicates that in the CA filaments orient in a
non-polar way~\cite{CCD+03}. It has been suggested 
that this holds for the PPB as well~\cite{TSG+04,DG03}.
Therefore, we neglect the polarization ${\bf p}$ and only keep the
nematic order ${\bf q}$. In situations where the total mass
of polymerized microtubules is 
conserved (such as the transition from CA to PPB~
\cite{Vos04}),  the source term $s$ can be neglected.
For simplicity, we focus 
on states of rotational symmetry around the cylinder axis and 
we neglect effects of the surface curvature. 

The cylinder axis is the $x$-axis of our coordinate frame, 
the second dimension described by
the $y$-axis is wrapped around the cylinder with radius $R$, see Fig.~\ref{fig:fig1}(c).
The dynamics of configurations which are rotationally invariant,
obey a set of equations projected on the cylinder axis, 
which can be derived by symmetry arguments \cite{LK01,KZJ03}:
\begin{eqnarray}
\eta ^{-1}\sigma _{xx} &=&A_{1}c+A_{2}c^{2}+A_{3}c\partial
_{x}^{2}c+A_{4}(\partial _{x}c)^{2}+A_{5}q_{xx}  \label{eq:evolprojc} \\
\omega _{xx} &=&-B_{1}q_{xx}+B_{2}\partial _{x}^{2}q_{xx}+B_{3}\partial
_{x}^{4}q_{xx}+B_{4}\partial _{x}^{2}c  \label{eq:evolprojqxx} \\
&&+B_{5}(q_{xx}^{2}+q_{xy}^{2})q_{xx}  \notag \\
\omega _{xy} &=&-B_{1}q_{xy}+B_{2}\partial _{x}^{2}q_{xy}+B_{3}\partial
_{x}^{4}q_{xy}  \label{eq:evolprojqxy} \\
&&+B_{5}(q_{xx}^{2}+q_{xy}^{2})q_{xy}  \notag
\end{eqnarray}
where we have neglected higher order terms. 
It has been shown that the terms in Eq. (\ref{eq:evolprojc})-(\ref{eq:evolprojqxy})  
arise from coarse-grained microscopic models and can 
be related to filament sliding induced by motor proteins 
in an immobile viscous background \cite{KJ00,KZJ03,LM03,tanaUP}. 
An example is the contractile tension given by the term $A_{2}c^{2}$
which has been discussed in \cite{KZJ03}.

Dimensional analysis reveals that these equations contain six independent
parameters. The values of only three of these turn out to  influence the
dynamics qualitatively; they can be identified with $A_{2}$, $B_{2}$, and 
$F=-A_5B_4$. The other 
parameters describe nonlinear terms which are required to stabilize the dynamics.
Stability requires $A_{3}>0$, $A_{4}<0$, $B_{3}<0$, and $B_{5}<0$. The specific values of these parameters do not qualitatively influence the asymptotic dynamics.

In order to get some general insight into this problem, we first consider
the somewhat artificial case of periodic boundary conditions in
$x$-direction with period $L$. In this case, 
the homogenous isotropic state, i.e., $c(x)=c_{0}$, 
$q_{xx}(x)=q_{xy}(x)=0$ for all $x$, is a stationary solution of the dynamic
equations (\ref{eq:evolprojc})-(\ref{eq:evolprojqxy}).
We first perform a linear stability analysis of this state.
The linearized equation for the perturbation
 $\delta q_{xy}$ decouples from the
corresponding equations for the perturbations $\delta c$ and $\delta q_{xx}$. 
The field $q_{xy}$ becomes unstable for $B_{1}<0$. 
This instability corresponds to the isotropic-nematic transition. 
Here, we focus on the instabilities of the isotropic homogenous 
state with $B_1>0$.

\begin{figure}
\includegraphics[scale=0.3]{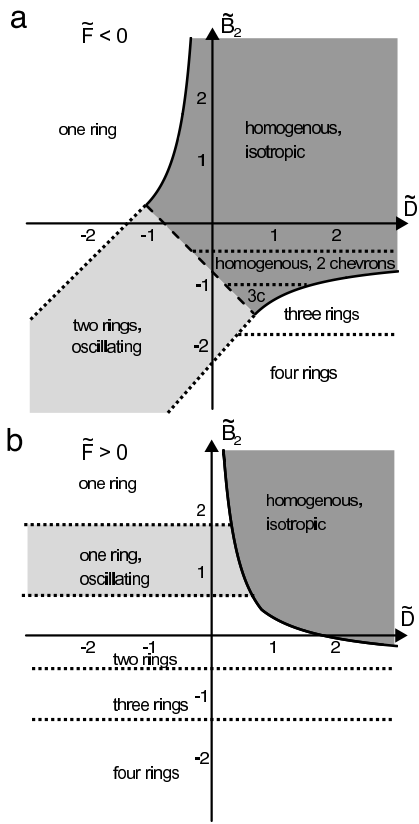}
\caption{\label{fig:phasespace}
Schematic state diagrams of an active gel layer in cylindrical geometry
with periodic boundary conditions in the $x$-direction,
are shown for two values of $\tilde F$
as a function of two
dimensionless parameters  $\tilde B_{2}$ and $\tilde D$.
Regions of linear stability of 
homogeneous filament distributions are shaded 
in dark grey. Outside these regions, the topology of the state diagram as  determined numerically is indicated by dotted lines. 
Asymptotic states include chevron patterns and single or multiple 
rings that are either stationary or do oscillate. 
Regions of oscillating solutions are shaded in light grey.
The long dashed line indicates a Hopf bifurcation determined by linear stability analysis. 
Examples for stationary chevron and ring patterns are displayed in Fig.~\ref{fig:gallery}.
Parameters are defined as $\tilde D =-(\tilde A_{1}+2\tilde c_{0}\tilde A_{2})$,
$\tilde B_2=B_2/(B_1 L^2)$, $\tilde F=F/(B_1^2L^4)$ and $\tilde A_{2}=A_{2}/(B_1 L^4)$.
Parameters values are
$\tilde c_{0}=c_{0}L=0.5$, $L(-2B_1/A_1)^{1/2}=5$,
$\tilde A_1=A_{1}/(B_1L^2)=-2$, 
$A_{3}/(B_1L^6)=0.1$, $A_{4}/(B_1L^6)=-1$, $A_5/B_1=1$,
$B_{3}/(B_1L^2)=-0.05$, $B_5/B_1=-10$ and $\tilde F=-1$ for (a) and $\tilde F=1$ for (b).
$L$ is the period of the system.} 
\end{figure}

\begin{figure}
\includegraphics[scale=0.46]{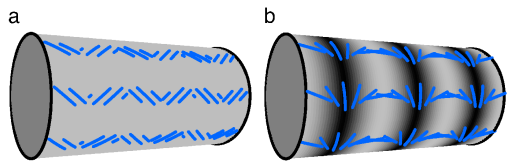}
\caption{\label{fig:gallery}
(color online)
Examples of stationary solutions to the dynamic equations 
with periodic boundary conditions.
(a) Chevron pattern with homogeneous filament density for $\tilde B_2=-1.25$, $\tilde D=0.75$, 
and $\tilde F=-1$.
(b) Multiple ring pattern for $\tilde B_2=-1.5$, $\tilde D=1.5$, 
and $\tilde F=-1$.}
\end{figure}

The linearized dynamics for $\delta c=\sum_{n}c_{k}\mathrm{e}^{2\pi inx/L}$
and $\delta q_{xx}=\sum_{n}q_{xx,k}\mathrm{e}^{2\pi inx/L}$ is given by
\begin{eqnarray}
\lefteqn{\frac{d}{dt}\left(
\begin{array}{c}
c_{k} \\
q_{xx,k}
\end{array}
\right) =}   \label{eq:evollin} \\
&&\left(
\begin{array}{cc}
-Dk^{2}-A_{3}c_{0}k^{4} & +A_{5}k^{2} \\
-B_{4}k^{2}/2 & -B_{1}-B_{2}k^{2}+B_{3}k^{4}
\end{array}
\right) \left(
\begin{array}{c}
c_{k} \\
q_{xx,k}
\end{array}
\right)  \notag \quad ,
\end{eqnarray}
where $D=-(A_{1}+2c_{0}A_{2})$ 
is an effective diffusion constant. Figure~\ref{fig:phasespace}
represents the region of stability of the homogenous isotropic state as a dark grey
area for sufficiently large values of the parameters $\tilde D$ and $\tilde B_2$.
The parameter $-\tilde D$ is related to the contractile tension in the system.
For sufficiently small $\tilde D$, the system tends to undergo density instabilities
which lead to ring solutions.
The parameter $\tilde B_2$ characterizes a length scale related to the orientational order.
For sufficiently small $\tilde B_2$ orientational instabilities occur.
The corresponding unstable mode consists of
periodically alternating regions with nematic order at an angle of $\pi /4$
and $-\pi /4$ with the $x$-axis. These structures are reminiscent of
chevrons in nematic liquid crystals, cf. Fig.~\ref{fig:gallery}(a) \cite{degennes}.
For $F<0$, the homogeneous state becomes unstable by either stationary instabilities
or by Hopf bifurcations which lead to traveling wave solutions. 
Because of symmetry, these
waves can propagate along the $x$-axis in both directions. 
Hopf bifurcations occur if $(D-B_2)^2+4F<0$. 
For $F>0$ instabilities towards inhomogeneous states are
always stationary.
The 
instabilities towards stationary states 
lead to periodic patterns of filament accumulation
combined with nematic order, parallel or perpendicular to the $x$-axis,
at the maxima of the filament density, see Fig.~\ref{fig:gallery}(b).
For decreasing $\tilde B_2$ the number of rings increases as the corresponding
length scale decreases.
In order to obtain the full phase diagrams displayed in Fig. 2, we solved the
dynamic equations (3)-(5) numerically with periodic boundary conditions\cite{fn}.

Periodic boundary conditions are inappropriate to describe 
cells. An important boundary condition is the zero flux condition 
$\bm{j}_{0,L}=0$, which implies that the cortical material 
cannot leave the cell. Additional boundary conditions have to be specified
to fully define the asymptotic solutions.
It is observed that the end-faces of interphase plant cells are essentially
free of microtubules. This is consistent with the absence of imposed order
at the boundaries and we therefor set $\bm{q}_{0,L}=0.$ 
In the following we use for simplicity the boundary conditions ${\bf j}=0$,
${\bf q}=0$ and  $\partial _{x}c=\partial _{x}q_{xx}=\partial_{x}q_{xy}=0$ 
at both $x=0$ and $x=L$. 

\begin{figure}
\includegraphics[scale=0.4,]{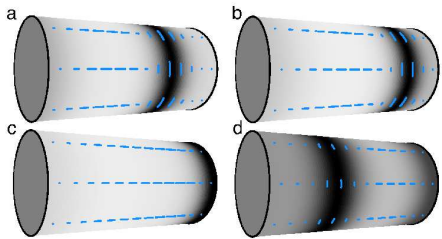}
\caption{\label{fig:oscillation}
(color online)
Snapshots at different times of an oscillating filament ring for no-flux boundary conditions
and $\tilde B_2=1$, $\tilde D=0.25$ and $\tilde F=1$. The pattern consists of a filament
ring which forms near the center of the cylinder (a) and moves towards the right pole (b,c), where the ring disappears. Simultaneously, a new ring is formed near the
center (d) which subsequently moves to the opposite pole. The whole process
is repeated periodically.
The times correspond to phases $\phi=\omega t$
of the oscillation with $\phi=0$ (a), $ \phi=(4/36) 2\pi$ (b), $ \phi=(7/36) 2\pi $ (c) and $ \phi=(15/36) 2\pi $(d). }
\end{figure}

Figure~\ref{fig:fig1}(c) shows a steady state solution of Eqs. (\ref{eq:evolprojc})-
(\ref{eq:evolprojqxy}) 
with zero flux boundary conditions.
This solution corresponds to a stationary filament ring with
filaments that are oriented in $y$-direction and localized in a ring-like
pattern wound around the cylinder, strongly reminiscent of the PPB. 
The corresponding stress is anisotropic
and for $\sigma_{yy} >\sigma_{xx}$ the ring solution is contractile.
Additional asymptotic solutions are similar to the solutions shown in Fig.
\ref{fig:gallery} with periodic boundary conditions.
Furthermore, oscillatory solutions appear along a line of Hopf- bifurcations. 
An example of such a state with no flux boundary conditions
is shown in  Fig.~\ref{fig:oscillation}.
The oscillating solutions for no flux boundary conditions correspond to 
traveling waves in the case of periodic boundary conditions. 
Additional states can be generated
via the boundaries 
by imposing other boundary conditions. 

In conclusion we have presented a generic approach to the dynamics of
a cortical layer of an active gel of cytoskeletal filaments near the cell
surface.
We have focussed on a situation which can describe the formation of
filament rings such as the PPB in plant cells and the
contractile ring in animal cells. 
Ringlike cytoskeletal patterns occur also in prokaryotes. An example is the
Z-ring formed by the tubulin analog  FtsZ \cite{lutk02}. 
We have analyzed the formation of patterns
via dynamic instabilities resulting from active processes such as the
action of motor proteins and the polymerization and depolymerization
of filaments. Of particular interest is the spontaneous formation of
filament rings which provides a possible mechanism for the formation 
of the PPB. Such ring structures
can be contractile and could induce constriction of a deformable cylinder.
Furthermore, we have found solutions corresponding to filament rings which
move periodically towards both ends of the cylinder. Oscillating actin
rings have been observed in lymphoblasts treated with nocodazole to remove
the microtubules~\cite{born89} 
and more recently in fragments of fibroblasts~\cite{ep05}. 
Qualitatively, they exhibit the same type of 
oscillatory pattern as described in Fig.~\ref{fig:oscillation}.

Our description assumes that filaments are
short as compared to the cylinder radius $R$. In the PPB, microtubules could
have a length of several tens of $\mu$m. In this case, the bending elasticity
can become important and could lead to anisotropies in our description
since azimuthal orientation would be energetically disfavored.
In order to connect our macroscopic approach with molecular properties
of filaments and associated proteins, more microscopic descriptions
are valuable. 
For example, negative values of the parameter $\tilde D$ in the phase diagram of
Fig. 2 can result from contractile stresses caused by the action of 
motor-aggregates~\cite{LM03,KZJ03}. 
Similarly, the parameter $\tilde B_2$ characterizes
a length scale associated with orientational order, which depends for example 
on filament lengths and packing density.
Such relations between microscopic models and phenomenological descriptions
could help to identify microscopic mechanisms, which underlie the formation of 
cortical rings in cells.

We thank Anne Mie Emons and Jan Vos for many stimulating discussions.
The work of MCL, CT, BMM and MD is part of the research program of
\textquotedblleft Stichting Fundamenteel Onderzoek der Materie
(FOM)\textquotedblright\ which is financially supported by \textquotedblleft
Nederlandse Organisatie voor Wetenschapelijke Onderzoek
(NWO)\textquotedblright .


\begin{thebibliography}{10}
\bibitem{fn} We used a forward Euler scheme to numerically integrate Eqs.~(1)-(5).
	 Space was discretized in steps of $\Delta x=0.01 L$. 
	 The time step was $\Delta t =10^{-7}/B_1$. 
\end{thebibliography}
\end{document}